\documentclass{article}
\usepackage[utf8]{inputenc}
\usepackage{amsmath}
\usepackage{amssymb}

\title{Martingale consumption}
\author{Peter Holm Nielsen \\ \footnotesize{PFA Pension, Sundkrogsgade 4, DK-2100 Copenhagen Ø, Denmark} \\ \footnotesize{E-mail: pei@pfa.dk}}
\date{\small May 26, 2025}

\usepackage{natbib}
\usepackage{graphicx}
\usepackage{verbatim}
\usepackage{latexsym}
\usepackage{ragged2e}

\newtheorem{proposition}{Proposition}

\newtheorem{lemma}{Lemma}
\newtheorem{remark}{Remark}

\newcommand{\condprob}[2]{\mathrm{P}\left(\left. #1 \, \right| #2 \right)}
\newcommand{\meanmeas}[2]{\mathrm{E}^{#2}\left( #1 \right)}
\newcommand{\condmean}[2]{\mathrm{E}\left(\left. #1 \, \right| #2 \right)}
\newcommand{\as}{\mathrm{a.s.}}

\newcommand{\be}{\begin{equation}}
\newcommand{\ee}{\end{equation}}
\newcommand{\qed}{\hfill $\Box$}

\begin{document}

\maketitle

\begin{abstract}
    We propose martingale consumption as a natural, desirable consumption pattern for any given (proportional) investment strategy. The idea is to always adjust current consumption so as to achieve level expected future consumption under the arbitrarily chosen investment strategy. This approach avoids the formulation of an optimization objective based on preferences towards risk, intertemporal consumption, habit formation etc. We identify general explicit solutions in deterministic-coefficient models. In the general case with random coefficients we establish uniqueness, but the question of existence of a solution is unsettled. With the interest rate as the only random factor we derive a PDE for the wealth-to-consumption factor as a function of the state variables, which, however, is non-linear and without known closed-form solutions. We briefly consider the discrete-time case and obtain similar results. Throughout, we compare with well-known optimal strategies for classical CRRA investors with time-additive utility of consumption and find that under suitable time preferences they may in certain cases achieve martingale consumption simultaneously.
\end{abstract}

\noindent \textbf{Keywords:} Consumption strategies; Martingales; PDE's.

\section{Introduction} 
Consumption strategies and patterns constitute a main topic of economics and (mathematical) finance. Often, they are studied along with investment strategies, forming combined investment/consumption problems in various financial market models.

The classical economical approach is that of utility optimization under certain assumptions on preferences towards risk, intertemporal substitution etc. This was first carried out in continuous-time models by \cite{Merton69, Merton71}, who derived explicit optimal investment and consumption strategies for agents with CRRA/HARA risk preferences and time-additive utility of consumption. Ever since, Merton's pioneering work has been generalized in numerous directions, e.g.\ in models with stochastic interest rates, e.g.\ \cite{DeelstraGrasselliKoehl00} or \cite{MunkSorensen04}, or general non-markovian financial market models via the martingale method, e.g.\ \cite{KaratzasLehoczkyShreve87} or \cite{CoxHuang89}, to name just a few.

The optimal consumption strategy derived by Merton in a ``standard'' model with constant coefficients constitutes a geometric Brownian motion, typically with strictly positive drift. This is perhaps somewhat at odds with a common-sense perspective on preferred consumption patterns for individuals (see below), as well as with observed consumption patterns.

Many papers aim to explain observed consumption patterns and/or to formulate various optimization objectives, where the assumption of time-additivity of utility of consumption is replaced by other assumptions so as to allow for, e.g., habit formation, where the current consumption level is measured in relation to past consumption levels in some way, such as in \cite{Munk08}, or other economically sensible features. A related branch of papers formulate objectives that aim more generally and perhaps more explicitly at consumption smoothing, where the goal is to obtain a more level and less volatile consumption stream, arguably more in line with preferred consumption patterns, e.g., \cite{BruhnSteffensen13} or \cite{ChoiJeonKoo22} and references therein. 

In this paper, we take a different and, to the author's knowledge, new approach regarding consumption strategies. Taking the (proportional) investment strategy as exogenously and a priori given, we aim to find a \textit{martingale} consumption strategy, for which we shall use the abbreviation MCS. If the consumption rate process forms a martingale, then the expected future consumption rate at any time, given the evolution until the present time, equals the current consumption rate. In particular, the expected future consumption rates all equal the initial rate at time zero. Thus, on the one hand, a certain kind of consumption smoothing takes place, as the consumption level is continuously adjusted to the current wealth so as to level out remaining expected future consumption, whereas on the other hand, the consumption level continuously moves in line with unanticipated investment gains and losses, fully allowing for paths resulting in non-smooth consumption due to good or bad investment scenarios. The idea is that the agent chooses the (proportional) investment strategy according to his individual preferences, which may (or may not) be inspired by results on optimal investment strategies, and then at any time accepts the current -- and expected future -- consumption rate that follows from the investment gains or losses, as long as the current rate equals the expected future rates, given the evolution. 

Our approach is motivated by a certain type of variable annuities employed in practice, i.e., unit-linked ``annuity'' pension products, where the proportional asset allocation of each retiree follows a predetermined scheme in the retirement phase, and the pension benefits are adjusted concurrently (in pratice, say, annually) to the wealth level resulting from unanticipated investment gains or losses and the updated expectations regarding future returns. This leads to ``market based annuities'', where the retirement benefits are allowed to fluctuate with market movements, but with annual readjustments in such a way that the \textit{expected} remaining benefits are always equal. 

Just to clarify the idea: It is well known that in a complete market, any consumption strategy satisfying the appropriate budget constraint of the market can be financed through a hedging investment strategy, so in particular, this also goes for any \textit{given} MCS. However, in this paper we take the (proportional) investment strategy as given and \textit{then} look for an MCS under that investment strategy. We let the consumption strategy depend on the investment strategy, not the other way around. 

As alluded to above, the underlying, ``implicitly-assumed'' desired consumption pattern of the agent is that of an annuity, which in its pure form has deterministic benefits of equal size, and thus in particular constitutes a martingale. Indeed, annuities have traditionally been employed broadly in pension schemes, and generally seem to be widely regarded as sound for retirees, for obvious reasons. However, an annuity imposes a conservative investment strategy in bonds. An MCS can be viewed as a generalization of an annuity that allows for investment strategies in more risky assets with higher expected returns and thus a higher expected consumption rate.

Our approach has links to certain aspects of some early research on consumption. In deterministic settings, \cite{ModiglianiBrumberg54} and \cite{Friedman57} formulated basic, sound principles for consumption versus saving over the life cycle. Their basic idea, also known as the permanent income hypothesis, was that a rational agent should seek to smooth out consumption over the entire life time by taking future income and future consumption needs into account in the decision of current consumption/savings rates, rather than, e.g., let current consumption be driven excessively by the current wage level and/or financial wealth (or changes therein). Along those lines, \cite{Hall78} showed that maximization of (expected) accumulated time-additive utility of future consumption in a discrete-time model with a fixed interest rate but, possibly, stochastic future wages, leads to martingale consumption if the agent's ``impatience'' parameter exactly offsets the interest rate. 

\cite{Hakansson69} solved the classical \textit{combined} consumption/investment problem in a discrete-time model and noted that ``the optimal consumption strategy satisfies the properties of the permanent (normal) income hypothesis''. This, however, is to be understood in the sense that the optimal consumption strategy is only a function of current wealth and the value of total future income. It also depends \textit{parametrically} on the agent's risk aversion and time preferences, and, just as in the continuous-time Merton model, the consumption process is thus not a martingale in general. However, as we shall see, a utility-optimizing agent may in some cases (although not all!) achieve martingale consumption \textit{simultaneously}, if desired, by a suitable choice of time preferences. 

In another related paper, \cite{Fischer08} proves existence and provides construction methods for general discrete-time consumption processes, in particular martingale ones, in a model with a general money market account (but no other assets). 

The rest of the paper is organized as follows: In Section~\ref{Sec:SimpleModels}, which serves as an appetizer, we introduce the problem and provide solutions in some simple models, and we compare with the classical utility-optimization approach. Section~\ref{Sec:GeneralCase} deals with the problem in a general financial market model with stochastic coefficient processes. We obtain a general uniqueness result and generalize the solutions from the simple models in Section~\ref{Sec:SimpleModels} to a certain subclass of models. In Section~\ref{Sec:StochInterestCase} we treat the problem in the case with the interest rate as the only random market factor and derive a PDE for the solution, which, alas, has no known explicit solutions. For completeness we briefly treat the discrete-time case in Section~\ref{Sec:DiscreteTimeCase}, and Section~\ref{Sec:Conclusion} concludes with a summary of the main results of the paper.
 
\section{Martingale consumption in simple models} \label{Sec:SimpleModels}
We consider an economic agent with initial wealth $x_0 > 0$, who consumes and invests continuously over a fixed time horizon $[0,T]$ for some $T > 0$. The wealth and consumption processes are denoted by $X(\cdot)$ and $c(\cdot)$, respectively.

Different annuity-type expressions will appear, so to ease notation we shall let $B_f(t)$ denote the time $t$ value of a continuous-time unit annuity over $[t,T]$ evaluated at rate $f(\cdot)$ for some (integrable) function $f:[0,T] \rightarrow \mathbb{R}$, i.e.,
\begin{equation} \label{B}
    B_f(t) = \int_t^T e^{-\int_t^u f(s) \, ds} \, du, \; 0 \le t \le T.
\end{equation}
If $f(\cdot)$ equals the interest rate in a deterministic interest rate setting, or more generally, if $f(\cdot)$ equals the (time $t$)-forward rate curve in a stochastic interest rate setting where such a curve exists, then $B_f(t)$ is just the (time $t$)-market value of an annuity certain over $[t,T]$. In particular, if $f(\cdot)$ is constant, say $f(\cdot) = \alpha \in \mathbb{R}$, then
$$
B_f(t) = B_{\alpha}(t) = \frac{1-e^{-\alpha(T-t)}}{\alpha}.
$$
If $f$ is continuous, then $B_f$ satisfies the differential equation
\begin{equation} \label{B_diff_eq}
\frac{dB_f(t)}{dt} = f(t) B_f(t) - 1.
\end{equation}
This section serves as an appetizer, where we introduce the problem and its solutions in some simple cases. To ease the reading here, we defer a mathematically rigorous treatment of the problem and its setup to Section~\ref{Sec:GeneralCase}. 

First, consider the case with deterministic interest rate $r(\cdot)$ and no available risky assets. The wealth dynamics are in this case given simply by
$$
dX(t) = X(t) r(t) \, dt - c(t) \, dt, \; 0 < t < T.
$$
The only MCS (exhausting all wealth over $[0,T]$) in this case is the one with constant consumption rate given by
\begin{equation} \label{c_deterministic}
c(t) = c(0) = \frac{X(0)}{B_{r(\cdot)}(0)}, \; 0 \le t \le T. 
\end{equation}
We compare this to the classical problem of optimization of utility of consumption, where an agent with constant relative risk aversion parameter $\gamma > 0$ and time preferences given by a continuous function $\beta : [0,T] \rightarrow \mathbb{R}$ has the objective 
$$
\max_c{\int_0^T e^{-\int_0^t \beta (s) \, ds} u(c(t)) \, dt},
$$
over all admissible consumption processes $c(\cdot)$, where $u: (0,\infty) \rightarrow \mathbb{R}$ is given by
$$
u(x) = \left\{ \begin{array}{ll}\frac{x^{1-\gamma}}{1-\gamma},& \mathrm{if } \; \gamma \neq 1, \\ \log(x), & \mathrm{if } \;\gamma = 1, \end{array} \right. 
$$
$x \in (0,\infty)$. It is well known, see e.g.\ \cite{Merton69}, that the optimal consumption stream is given in feedback form (on $[0,T)$) by
\begin{equation} \label{c_util_deterministic}
c(t) = \frac{X(t)}{B_{f_1(\cdot)} (t)}, \; 0 \le t < T,    
\end{equation}
where 
\begin{equation} \label{f_1}
f_1(t) = \frac{\beta(t) - (1-\gamma) r(t)}{\gamma}, \; 0 \le t \le T.    
\end{equation}
Simple differentiation in this case shows that 
\begin{equation} \label{c_util_deterministic_diff}
dc(t) = c(t) \frac{r(t)-\beta(t)}{\gamma} \, dt, \; 0 < t < T,    
\end{equation}
(from which $c(T)$ can naturally be defined as $c(T)=\lim_{t\rightarrow T}c(t)\in (0,\infty)$.)
Thus, whenever $r(t)>\beta(t)$, the optimal consumption rate is increasing. The growth rate is typically small, however, although it can be large for a low risk-averter (with $\gamma < 1$), who prefers to postpone consumption in order to gain interest and thus increase total consumption. 

If $\beta(\cdot) \equiv r(\cdot)$, the optimal consumption rate for \textit{any} utility-optimizing agent will be constant and obviously equal to the constant consumption level in (\ref{c_deterministic}), and in particular a martingale (this is the continuous-time analogue of \cite{Hall78}'s result mentioned in the introduction). Thus, a utility-optimizing agent can achieve martingale consumption \textit{simultaneously} if desired; this would imply a time preference function equal to the deterministic interest rate. If $ \beta(\cdot) \not\equiv r(\cdot)$, then $c$ will not be a martingale.

Now, if the agent can also invest in a risky asset with price dynamics given by
$$
dS(t) = S(t)\left(r(t)+\lambda \sigma \right) \, dt + S(t) \sigma \, dW(t),
$$
where $\lambda,\sigma \in \mathbb{R}$, the ``classical'' problem of (\cite{Merton69}) is to optimize the investment \textit{and} consumption processes, i.e.,
\be \label{eq:ClassicalUtilityOptimization}
\max_{(c,\pi)}{\int_0^T e^{-\int_0^t \beta (s)} u(c(t)) \, dt},
\ee
over all admissible investment-consumption processes $(\pi(\cdot),c(\cdot))$, still with the objective of maximizing expected accumulated utility of consumption under the specified time preferences. Here, $\pi(\cdot)$ denotes the proportion of wealth invested in the risky asset (well defined on $[0,T)$), yielding the wealth dynamics 
\begin{equation} \label{Wealth_dynamics_BS}
dX(t) = X(t) \left(r(t) + \pi(t) \lambda \sigma \right) \, dt + X(t) \pi(t) \sigma \, dW(t) - c(t) \, dt, \; 0 < t < T. 
\end{equation}
Again, the solution in this case is well known. The optimal investment strategy is given explicitly by  
\begin{equation} \label{pi_Merton}
\pi(t) = \frac{\lambda}{\sigma \gamma}, \, 0 \le t \le T,     
\end{equation}
and the optimal consumption stream is given in feedback form (on $[0,T)$) by 
\begin{equation} \label{c_util_determ_stock}
c(t) = \frac{X(t)}{B_{f_2(\cdot)} (t)}, \; 0 \le t < T,    
\end{equation}
where 
\begin{equation} \label{f_2}
f_2(t) = f_1(t)+(\gamma-1)\lambda^2/(2\gamma^2), \; 0 \le t \le T.    
\end{equation}
(Again, $c(T)=\lim_{t\rightarrow T}c(t)\in (0,\infty)$). 

Comparing $f_2$ to $f_1$, we see that the utility-optimizing agent, given the extra opportunity of investing in a risky asset, adjusts the \textit{wealth-to-consumption} factor $B$ by adding the term $(\gamma-1)\lambda^2/(2\gamma^2)$ to the rate determining the factor. Thus, with investments in a risky asset, a high risk averter (with $\gamma > 1$) will consume more (at the same wealth level), letting the expected future risk premium from the risky asset effect current consumption positively. A low risk averter (with $\gamma < 1$), on the other hand, will actually consume less and thus try to increase his current wealth in order to gain even larger future returns and thus increase future consumption. The log-investor (with $\gamma =1$) is, as usual, indifferent and does not adjust the wealth-to-consumption factor.  

The dynamics of $c(\cdot)$ are in this case 
\begin{equation} \label{c_util_determ_stock_diff}
dc(t) = c(t) \left(r(t)-\beta(t) +\frac{\lambda^2(\gamma + 1)}{2\gamma}  \right)\frac{1}{\gamma} \, dt + c(t) \frac{\lambda}{\gamma} \, dW(t), \; 0 < t < T.    
\end{equation}
The drift term of $c(\cdot)$ is decreasing in $\gamma$, so high risk averters will choose smoother consumption streams than low risk averters. If $r(\cdot) \equiv \beta(\cdot)$, then $c(\cdot)$ is a geometric Brownian motion. It is a  martingale if (and only if) 
\be \label{eq:BetaRequirementMG}
\beta(t) = r(t)+\lambda^2(\gamma + 1)/(2 \gamma),
\ee
for Lebesgue-a.e.\ $t \in [0,T]$. Thus, also in this case, any utility-optimizing agent can achieve martingale consumption simultaneously if desired, with the implicit time preference function given by (\ref{eq:BetaRequirementMG}). One would perhaps normally expect to have $\beta(\cdot) < r(\cdot)+\lambda^2(\gamma + 1)/(2 \gamma)$, which would make $c(\cdot)$ a strict submartingale. 

%From (\ref{pi_Merton}) we see that the optimal proportion in the risky asset is constant and thus in particular independent of the wealth process $X(\cdot)$.

Returning now to the quest for martingale consumption in the case with a risky asset for an agent, whose investment strategy is not necessarily utility-optimizing, we consider consumption streams of the form 
\begin{equation} \label{c_form_BS}
c(t) = \frac{X(t)}{a(t)}, \; 0 \le t < T,    
\end{equation}
for some deterministic function $a \in C^1[0,T]$ with $a(t)>0, \; 0\le t < T$. Then $c$ has the dynamics
\begin{equation}
dc(t) = c(t) \left( r(t)+\pi(t)\lambda \sigma - \frac{1}{a(t)}\left(a'(t) +1\right)\right) \, dt 
+ c(t) \pi(t) \sigma \, dW(t). \label{c_dynamics_BS}
\end{equation}
Thus, for any \textit{deterministic} investment strategy $\pi(\cdot)$ we can find an MCS given by (\ref{c_form_BS}) by letting $a$ be the solution to the ODE 
\begin{equation} \label{a_ODE_BS}
0 = a'(t) +1 -a(t)\left( r(t)+\pi(t)\lambda \sigma\right)    
\end{equation}
on $(0,T)$ with boundary condition $a(T) =0$ (to be motivated later). Using (\ref{B_diff_eq}) we easily obtain the solution $a=B_{f_3}$ with 
\begin{equation} \label{f_3}
f_3(t) = r(t) +\pi(t)\lambda \sigma, \; 0 \le t < T.    
\end{equation}
We note that $f_3(t) \, dt$ expresses the expected investment return in a short time interval $(t,t+dt)$. The solution thus naturally generalizes the one from the case without risky assets, see (\ref{c_deterministic}). 

In particular, with the constant-proportion investment strategy (\ref{pi_Merton}), which is optimal in the utility-optimization case, the diffusion terms of the consumption processes for the utility-optimizer and the martingale consumer ((\ref{c_util_determ_stock_diff}) and (\ref{c_dynamics_BS})) are the same, so the difference lies only in the drift terms (and, of course, in the initial consumption levels). As noted above, the drift term in (\ref{c_util_determ_stock_diff}) vanishes if $\beta(\cdot)$ is chosen so as to fulfill (\ref{eq:BetaRequirementMG}), making the two consumption processes coincide. Otherwise, the utility-optimizing agent will typically suffice with a lower initial consumption level than that of an agent who wants martingale consumption, in order to gain a higher future consumption level. 

\section{The general case} \label{Sec:GeneralCase}
In this section we formalize and analyze the problem in a fairly general, Brownian model of a complete financial market. We refer to \cite{KaratzasShreve98} for background material and details. We (still) consider an economic agent endowed with initial wealth $x_0>0$ and a planning horizon $T>0$. The investment opportunities consist of a locally risk-free money market account with price process $S_0$ and $n \ge 1$ risky assets with price processes $S_1,\ldots,S_n$, which can be thought of as, e.g., stocks or (zero coupon) bonds. 

All randomness is driven by a standard, $n$-dimensional Brownian motion, $W=(W_1,\ldots,W_n)^T$, defined on some probability space $(\Omega,\mathcal{F},\mathrm{P})$. We define the filtration $(\mathcal{F}_t)_{t \in [0,T]}$ by $\mathcal{F}_t = \sigma(\mathcal{F}^W_t \cap \mathcal{N}), \; 0 \le t \le T$, where $\mathcal{N}$ is the set of all subsets of $\mathcal{F}$ with $\mathrm{P}$-measure zero. 

The price processes are assumed to be strictly positive and have the following dynamics:
\begin{align*}
    dS_0(t)/S_0(t) & = r(t) \, dt, \\
    dS_i(t)/S_i(t) & = \alpha_i(t) \, dt + \sigma_i(t) \, dW(t), \; i =1,\ldots,n,
\end{align*}
where $r(\cdot)$ is the short interest rate, and $\alpha_i$ and $\sigma_i = (\sigma_{ij})_{j=1,\ldots,n}$ are the drift and volatility coefficients, respectively, of the $i$'th risky asset. These are all modeled as progressively measurable processes satisfying $\int_0^T | r(t) | \, dt < \infty, \; \as$, and $\int_0^T | \alpha_i(t) | + \sigma_{ij}^2(t) \, dt < \infty, \; \as$% and $\int_0^T\sigma_{ij}^2(t) \, dt < \infty \; \as$ 

We assume that $\sigma(t)$ is nonsingular for Lebesgue-a.e.\ $t \in [0,T]$, a.s. The process of \textit{market prices of risk}, $\lambda=(\lambda_1,\ldots,\lambda_n)^T$ is then defined by   
$$
\lambda(t) = \sigma^{-1}(t) (\alpha(t)-\underline{1}r(t)), 
$$
for Lebesgue-a.e.\ $t \in [0,T]$, a.s., where $\underline{1} \in \mathbb{R}^n$ is the unit (column) vector. We also assume that $\int_0^T \| \lambda(t)\|^2 \, dt < \infty, \; \as$, and that $\mathrm{E}(\Lambda(T)) = 1$, where 
$$
\Lambda(T) = e^{-\int_0^T \lambda(t) \, dW(t) -\frac{1}{2} \int_0^T \lambda^2(t) \, dt},
$$
so that $\mathrm{Q}$ defined by $d\mathrm{Q} = \Lambda(T) \, d\mathrm{P}$ is a martingale measure for the market, i.e., $W^\mathrm{Q}$ given by $W^\mathrm{Q}(t) = W(t) + \int_0^t \lambda(s) \, ds, \; t \ge 0,$ is a standard ($n$-dimensional) Brownian motion under $\mathrm{Q}$. As noted, the market is complete.

The agent is assumed to invest and consume continuously. We denote by $c(t)$ and $\pi_i(t)$ the consumption rate and the \textit{fraction} of wealth invested in the $i$'th risky asset, respectively, at time $t \in [0,T]$, $i=1,\ldots,n$. The fraction of wealth invested in the money market account is $1-\sum_{i=1}^n\pi_i(t)$. We shall only allow for investment/consumption strategies for which the wealth remains nonnegative throughout $[0,T]$. Formally, a consumption strategy is a progressively measurable, nonnegative RCLL process $c$ satisfying $\int_0^T c(t) < \infty, \; \as$, and an investment strategy is a measurable, $(\mathcal{F}_t)$-adapted $n$-dimensional process $\pi=(\pi_1,\ldots,\pi_n)$ satisfying $\int_0^T |\pi(t)\alpha(t)| + \|\pi(t)\sigma(t)\|^2 \, dt < \infty, \; \as$ (note that here, we differ from the notation in \cite{KaratzasShreve98}, since they generally let $\pi$ denote the \textit{amounts} invested in the assets).

The wealth process evolves according to the SDE
\begin{align}
    X(0) & = x_0, \nonumber \\ 
    dX(t) & = X(t) \, (r(t)+\pi(t) \sigma(t) \lambda(t)) \, dt + X(t) \pi(t) \sigma(t) \, dW(t) - c(t) \, dt. \label{Wealth_dynamics_general}
\end{align}
We can write (\ref{Wealth_dynamics_general}) in terms of $W^\mathrm{Q}$ as
\be \label{Wealth_dynamics_general_Q}
   dX(t) = X(t) r(t) \, dt + X(t) \pi(t) \sigma(t) \, dW^\mathrm{Q}(t) - c(t) \, dt.
\ee
It is well known that the SDE (\ref{Wealth_dynamics_general_Q}) for the wealth $X$ has a unique solution given by
\be \label{Wealth_general}
X(t) = Y^\pi(t) x_0 - Y^\pi(t) \int_0^t \frac{c(s)}{Y^\pi(s)} \, ds, \; 0 \le t \le T,
\ee
where, for $t \in [0,T]$,
\begin{align*}
Y^\pi(t) &= e^{\int_0^t r(s) \, ds}  \zeta^\pi(t) = S_0(t)\zeta^\pi(t),\\
\zeta^\pi(t) &= e^{\int_0^t \pi(s)\sigma(s) \, dW^\mathrm{Q}(s) -\frac{1}{2}\int_0^t \|\pi(s)\sigma(s)\|^2 \, ds}.
\end{align*}
For clarity, we use the superscript $\pi$ on $Y^\pi$ and $\zeta^\pi$ to make it clear that they depend on the investment strategy $\pi$. 

\begin{remark}
    %\normalfont 
    Since the wealth process is not allowed to become negative, it is not a restriction to model the investment strategy through the \textit{fractions} of wealth allocated in the risky assets: If the investment strategy had been modeled through the \textit{amounts} invested in the risky assets, the wealth would stay at zero if it ever became zero before $T$, and no more investment (or consumption) would take place (see \cite{KaratzasShreve98}, Remark~3.3.4). \qed % \hfill $\Box$
\end{remark}

Now, we are interested in MCS's under a \textit{given} investment strategy $\pi$. We impose the requirement that the wealth (including financial gains) be fully exhausted at time $T$, i.e.,
\begin{equation} \label{Wealth_exhaustion}
    X(T) = 0, \; \mathrm{a.s.}
\end{equation}
This requirement is motivated by the problem as such: We are looking for consumption strategies that are martingales over $[0,T]$ and for which all wealth is consumed over $[0,T]$. However, (\ref{Wealth_exhaustion}) will not be ``automatically'' fulfilled in our setup unless explicitly required: If $c$ is an MCS satisfying (\ref{Wealth_exhaustion}), then for any $k \in (0,1)$, the scaled consumption process $kc \equiv (kc(t))_{t\in [0,T]}$ is also a martingale, but $kc$ does not satisfy (\ref{Wealth_exhaustion}). The necessity of this condition in our setup is in contrast to consumption \textit{optimization} problems, where it should (hopefully) be automatically fulfilled by the optimization criterion.

Note that from (\ref{Wealth_general}) and the fact that $Y^\pi(T) > 0$, we see that (\ref{Wealth_exhaustion}) is equivalent to 
\be \label{Wealth_exhaustion_with_c}
\int_0^T \frac{c(t)}{Y^\pi(t)} \, dt = x_0, \; \as
\ee
We shall also require that 
\be \label{C_StrictlyPositive} 
c(t)>0, \, \forall t \in [0,T], \, \as
\ee
This requirement is primarily for technical reasons (see Remark~\ref{Rem:CSpecialForm} below), but it is economically reasonable, since we want smooth consumption over the entire time horizon. For a given investment strategy $\pi$ satisfying the technical assumptions above, we say that a consumption strategy $c$ is $\pi$-admissible if it satisfies (\ref{Wealth_exhaustion_with_c}) and (\ref{C_StrictlyPositive}) (and the technical assumptions above). 

An important fact for our problem is that any RCLL martingale is continuous in our setup (\cite{KaratzasShreve91}, Problem~3.4.16). Thus, we need only consider continuous consumption strategies. This is also quite natural, given our quest for martingale consumption.

We now have the following result on uniqueness. 
\begin{proposition}
    Let the investment strategy $\pi$ be given. If $c$ is a $\pi$-admissible consumption strategy and a martingale, then it is unique in the sense that if $\tilde{c}$ is also a $\pi$-admissible consumption strategy and a martingale, then $c$ and $\tilde{c}$ are indistinguishable, i.e., $\mathrm{P}(c(t) = \tilde{c}(t), \; \forall t \in [0,T]) = 1$. 
\end{proposition}
\textsc{Proof}. Let $c$ and $\tilde{c}$ be two $\pi$-admissible consumption strategies that are both martingales. Then $c-\tilde{c}$ is a continuous martingale, and by (\ref{Wealth_exhaustion_with_c}),
$$
\int_0^T \frac{(c-\tilde{c})(t)}{Y^\pi(t)} \, dt = 0, \; \as
$$
The assertion now follows from the following general lemma. 
%\hfill $\Box$
\qed

\vspace{1em}

\begin{lemma}
    Let $M$ be a continuous martingale on $[0,T]$, and let $A$ be a continuous, strictly positive (progressively measurable) process, (both defined on $(\Omega,\mathcal{F},(\mathcal{F}_t)_{t\in[0,T]},\mathrm{P})$) such that $\int_0^T |A(s)M(s)| ds < \infty, \; \as$ If $\int_0^T A(s)M(s) ds = 0, \; \as$, then $\mathrm{P}(M(t) = 0, \; \forall t \in [0,T]) = 1$. 
\end{lemma}
\textsc{Proof}. For notational convenience, let $H(t) = \int_0^t A(s)M(s) ds, \; 0 \le t \le T$. Note that $H \in C^1[0,T], \; \as$ For $\epsilon > 0$, and with $\inf \emptyset = T$, define the random times
\begin{align*}
\tau^\epsilon_1 &= \inf{ \left\{t \in [0,T]: |H(t)| \ge \epsilon \right\} }, \\
\tau^\epsilon_2 &= \inf{ \left\{t \in (\tau^\epsilon_1,T]: | H(t)-H(\tau^\epsilon_1)| > \epsilon/2 \; \; \mathrm{and} 
 \; \; H(\tau^\epsilon_1)(H(t)-H(\tau^\epsilon_1)) < 0 \right\} }.
\end{align*}
Note that $\tau^\epsilon_1$ is a stopping time and $\tau^\epsilon_2$ is an optional time. We prove the assertion by contradiction, so assume that $\mathrm{P}(M(t) = 0, \; \forall t \in [0,T]) < 1$. Then there exists an $\epsilon > 0$ such that $\mathrm{P}(\tau^\epsilon_1 < T) > 0$. On $(\tau^\epsilon_1 < T)$ we have either (i) $H(\tau^\epsilon_1)=\epsilon$ or (ii) $H(\tau^\epsilon_1)=-\epsilon$. 

On $G^+:= (\tau^\epsilon_1 < T) \cap (H(\tau^\epsilon_1)=\epsilon)$ we must have $H'(\tau^\epsilon_1) = A(\tau^\epsilon_1)M(\tau^\epsilon_1) \ge 0$ and thus $M(\tau^\epsilon_1) \ge 0$ ($\as$). Since $H(T) = 0, \; \as$, we must also have 
$$
\condprob{H(T)-H(\tau^\epsilon_1) = -\epsilon}{\mathcal{F}_{\tau^\epsilon_1}}=1,
$$ 
$\as$ on $G^+$. But this means in particular that $\condprob{\tau^\epsilon_2 < T}{\mathcal{F}_{\tau^\epsilon_1}}=1, \, \as$ on $G^+$. Now, on $G^+ \cap (\tau^\epsilon_2 < T)$, we must have $H'(\tau^\epsilon_2)=A(\tau^\epsilon_2)M(\tau^\epsilon_2)<0$ and in particular $M(\tau^\epsilon_2)<0$ ($\as$), since $H(\tau^\epsilon_2)-H(\tau^\epsilon_1)=-\epsilon/2$, and $H(\tau^\epsilon_2+x)-H(\tau^\epsilon_1)<-\epsilon/2, \; \forall x \in (0,h),$ for some $h > 0$ ($\as$). By the Optional Sampling Theorem (e.g., \cite{KaratzasShreve91}, Theorem~1.3.22), we have $\condmean{M(\tau^\epsilon_2)}{\mathcal{F}_{\tau^\epsilon_1}} = M(\tau^\epsilon_2), \; \as$, but this equality is violated ($\as$) on $G^+ \in \mathcal{F}_{\tau^\epsilon_1}$. Similar arguments show that it is also violated ($\as$) on $G^-:= (\tau^\epsilon_1 < T) \cap (H(\tau^\epsilon_1)=-\epsilon)$. Since $\mathrm{P}(G^+)+\mathrm{P}(G^-)= \mathrm{P}(\tau^\epsilon_1 < T) > 0$, we have a clear contradiction, so we conclude that $\mathrm{P}(M(t) = 0, \; \forall t \in [0,T]) = 1$. \qed%\hfill $\Box$

\vspace{1em}

\noindent
In the following we shall in particular consider consumption processes that on $[0,T)$ are of the form 
\be \label{c_defined_by_a} 
    c(t) = X(t)/Z(t), \; 0 \le t < T,
\ee
for some strictly positive, continuous process $Z=(Z(t))_{t\in[0,T)}$ of the form 
\begin{align}
    Z(0) & = z_0, \nonumber \\ 
    dZ(t) & = \alpha_Z(t) \, dt + \sigma_Z(t) \, dW(t), \label{Z_dynamics}
\end{align}
for some $z_0 > 0$ and some progressively measurable processes $\alpha_Z$ and $\sigma_Z=(\sigma_{Z1},\ldots,\sigma_{Zd})$ satisfying $\int_0^T |\alpha_Z(t)| + \|\sigma_Z(t)\|^2 \, dt < \infty, \as$, such that 
\be \label{Z_Integral_cond}
\int_0^T1/Z(t) \, dt = \infty, \; \as
\ee
The condition (\ref{Z_Integral_cond}) implies in particular that $\lim_{t\rightarrow T}Z(t) = 0, \; \as$, and is motivated below. We shall refer to a process $Z$ satisfying these conditions as an admissible \textit{wealth-to-consumption factor} process. 

\begin{remark} \label{Rem:CSpecialForm}
%\normalfont 
Since we are looking for MCS's, it is not a restriction only to consider consumption processes of the form (\ref{c_defined_by_a}). To see this, note that by the (local) martingale representation theorem (e.g.\ \cite{KaratzasShreve91}, Problem~3.4.16) we know that if $c$ is a local martingale, it has the form $c(t) = c(0)+\int_0^t \theta(s) \, dW(s)$ for some progressively measurable $n$-dimensional process $\theta$. Since $c$ must be strictly positive (recall (\ref{C_StrictlyPositive})), we can \textit{define} $Z$ by $Z(\cdot) = X(\cdot)/c(\cdot)$ so that (\ref{c_defined_by_a}) is fulfilled. By Ito's formula, $Z$ then has the form (\ref{Z_dynamics}). \qed %\hfill $\Box$
\end{remark}

With $c$ as in (\ref{c_defined_by_a}), the SDE (\ref{Wealth_dynamics_general}) for $X$ on $[0,T)$ becomes
\begin{align}
    X(0) & = x_0, \nonumber \\ 
    dX(t) & = X(t) \, \left(r(t)-\frac{1}{Z(t)}+\pi(t) \sigma(t) \lambda(t)\right) \, dt + X(t) \pi(t) \sigma(t) \, dW(t), \label{X_dynamics_with_c_as_factor}
\end{align}
which has the solution
$$
X(t) = x_0 \exp \left(\int_0^t r(s)-\frac{1}{Z(s)} \, ds\right) \zeta^\pi(t), \; 0 \le t < T.
$$
From the conditions above it is seen that $\lim_{t\rightarrow T}X(t) = 0, \, \as$, so we may naturally define $X(T) := \lim_{t\rightarrow T}X(t) = 0, \, \as$, and in particular, we note that (\ref{Wealth_exhaustion}) is satisfied, which is the motivation for (\ref{Z_Integral_cond}). 

The goal now is to find an admissible wealth-to-consumption factor process $Z$ so that $c$ becomes a martingale. On $[0,T)$ we have from Ito's formula that
\begin{equation} 
dc(t) = \frac{dX(t)}{Z(t)}-\frac{(X(t)+dX(t))dZ(t)}{Z^2(t)}+\frac{X(t)(dZ(t))^2}{Z^3(t)}.    
\end{equation}
Using (\ref{c_defined_by_a}), (\ref{Z_dynamics}), and (\ref{X_dynamics_with_c_as_factor}), and performing a few lines of algebra, we obtain
\begin{align}
dc(t) &= c(t)\left(r(t)-\frac{1}{Z(t)}+\pi(t) \sigma(t) \lambda(t)\right) \, dt + c(t) \pi(t) \sigma(t) \, dW(t) \nonumber \\
&  - \frac{c(t)}{Z(t)} \left(\pi(t) \sigma(t) \sigma^T_Z(t) + \alpha_Z(t) - \frac{\|\sigma_Z(t)\|^2}{Z(t)}\right)\, dt - \frac{c(t)}{Z(t)} \sigma_Z(t) \, dW(t).  \label{c_dynamics_general}  
\end{align}
Now, for $c$ to be a martingale (or just a local martingale) we need the drift term to be zero, which requires that for every $t\in [0,T)$, $\as$,
\be \label{c_drift_zero_general}
0 = \left(r(t)+\pi(t) \sigma(t) \lambda(t)\right)- \frac{1}{Z(t)} \left(1+\pi(t) \sigma(t) \sigma^T_Z(t) + \alpha_Z(t)\right) + \frac{\|\sigma_Z(t)\|^2}{Z^2(t)}. 
\ee
Alas, this seems to be a dead end in the general case. We cannot provide solution processes $Z$ in general, let alone determine whether a solution exists. In particular, the solution from the case with a single risky asset, deterministic coefficients and deterministic investment strategy, provided in Section~\ref{Sec:SimpleModels}, with the expected infinitesimal return as the ``annuity rate'' (see (\ref{f_3})), does \textit{not} have a straightforward generalization. To get an MCS in the general case it is not enough to consider the expected infinitesimal return; one also has to take the effects of changes in the market coefficients into account. This is akin to results on optimal investment strategies in utility optimization problems, where it is well known that the optimal asset allocation (for non-log-investors) contains some degree of hedging against adverse changes in the random market coefficients.

However, the dead end contains a small opening. If $r$, $\alpha$, and $\pi$ are deterministic, the result presented in Section~\ref{Sec:SimpleModels} pertaining to the case with a single risky asset can in fact be generalized, even with stochastic volatility coefficients. To see this, note first that if $Z(t) = B_f(t), \; 0 \le t < T$, i.e., if $Z$ is deterministic, then $\sigma_Z \equiv \underline{0}$, and $\alpha_Z(t) = Z(t)f(t)-1$, so (\ref{c_dynamics_general}) becomes
\be \label{c_dynamics_with_deterministic_Z}
dc(t) = c(t)\left[r(t)+\pi(t) \sigma(t) \lambda(t)-f(t)\right] \, dt + c(t) \pi(t) \sigma(t) \, dW(t).
\ee
In particular, in this case, $\lim_{t\rightarrow T}c(T) \in [0,\infty), \; \as$, so $c(T)$ can be defined as $c(T) := \lim_{t\rightarrow T}c(T)$. We state this generalization as a proposition.
\begin{proposition} \label{PropSolutionWithDeterministicCoeffs}
    If $r$, $\alpha$, and $\pi$ are deterministic, then the consumption process given by $c(t)=X(t)/B_f(t), \; 0 \le t < T,$ (and $c(T) = \lim_{t \rightarrow T}c(t)$) with $f:[0,T]\rightarrow \mathbb{R}$ given by 
    $$
    f(t) = r(t)+\pi(t) \sigma(t)\lambda(t)=r(t)+\pi(t) (\alpha(t)-\underline{1}r(t)), \; 0 \le t \le T,
    $$
    is a local martingale satisfying (\ref{Wealth_exhaustion_with_c}), so that $X$ satisfies (\ref{Wealth_exhaustion}). It is, in particular, also a supermartingale.

    Moreover, if $\mathrm{E}(\frac{1}{2}\int_0^T \|\pi(t)\sigma(t)\|^2)<\infty$, then $c$ is a true martingale.
\end{proposition}
\textsc{Proof}. Assume that $r$, $\alpha$, and $\pi$ are deterministic, and define $c$ as stated in the proposition. We then have from (\ref{c_dynamics_with_deterministic_Z}) that on $[0,T)$,
\be \label{c_dynamics_general_with_some_deterministic_coefficients} 
dc(t) = c(t) \pi(t) \sigma(t) \, dW(t), 
\ee
so 
$$
c(t) = \frac{x_0}{B_f(0)} \exp \left(\int_0^t \pi(s) \sigma(s) \, dW(s)-\frac{1}{2}\int_0^t \|\pi(s)\sigma(s)\|^2 \, ds \right), \; 0 \le t \le T.
$$
In particular, $c(T) := \lim_{t \rightarrow T}c(t)$ is well-defined, and $c$ is a local martingale. Being nonnegative, it is in particular also a supermartingale. The assertion that $X$ satisfies (\ref{Wealth_exhaustion}) follows from the analysis above, since $\int_0^T 1/B_f(t) \, dt = \infty$. The last assertion follows from Novikov's condition.\qed
%\hfill $\Box$

\vspace{1em}
Although we cannot provide MCS's in the general case, we end this section on a positive note by considering, as in Section~\ref{Sec:SimpleModels}, the optimal consumption process under the classical expected utility optimization objective (\ref{eq:ClassicalUtilityOptimization}) in the general model of this section (with general random coefficients), which is well known (see, e.g., \cite{KaratzasShreve98}, Ch.~3). The optimal \textit{investment} strategy generally depends on the coefficient processes, but the dynamics of the optimal \textit{consumption} process given by (\ref{c_util_determ_stock_diff}) in the simple model with a single risky asset and deterministic coefficients carry over to the present case with straightforward generalizations, i.e., in the general case the dynamics of the optimal consumption process are given by
\be \label{c_util_general_case_diff}
dc(t) = c(t) \left(r(t)-\beta(t) +\frac{\|\lambda(t)\|^2(\gamma + 1)}{2\gamma}  \right)\frac{1}{\gamma} \, dt + c(t) \frac{\lambda^T(t)}{\gamma} \, dW(t), \; 0 < t < T,    
\ee
This holds even if $\beta(\cdot)$ is allowed to be stochastic (but progressively measurable), so any utility-optimizing agent can indeed simultaneously achieve martingale consumption if desired, with the implicit time preference process given by (\ref{eq:BetaRequirementMG}), with $\lambda^2$ replaced by $\|\lambda(t)\|^2$.

A stochastic time preference process of this form (with $\beta(\cdot)$ as a stochastic process) is perhaps hard to justify economically, so this finding is, in its general form, arguably interesting mostly as an existence result. However, if $r(\cdot)$ and $\lambda(\cdot)$ are deterministic, then so is $\beta(\cdot)$ when given by (\ref{eq:BetaRequirementMG}), so in this case we have an economically more sound result.

If $r(\cdot)$ (or $\lambda(\cdot)$) is stochastic, but $\beta(\cdot)$ is not allowed to be, then the utility-optimizer can no longer achieve martingale consumption simultaneously.

\section{Stochastic interest rate case} \label{Sec:StochInterestCase}
We now turn to a special case of the general model considered in Section~\ref{Sec:GeneralCase}, with a fairly simple Markovian structure of the market and the interest rate as a single random factor. We let $n=2$. The (short) interest rate is assumed to  have dynamics of the form 
\begin{equation} \label{short_rate_general}
    dr(t) = \mu(t,r(t)) \, dt - \sigma_r(t,r(t)) \, dW_1(t),
\end{equation}
where $\mu: [0,T] \times E \rightarrow \mathbb{R}$ and $\sigma_r : [0,T] \times E \rightarrow (0,\infty)$ are functions sufficiently regular to allow for a unique, nonexploding (possibly weak) solution to the SDE (\ref{short_rate_general}) with values in some open state space $E \subseteq \mathbb{R}$.

Moreover, $\sigma(t)$ is assumed to have the form 
$$
\sigma(t) = \left(\begin{array}{ll}
    \sigma_{11}(t,r(t)) & 0 \\
     \sigma_{21}(t,r(t)) & \sigma_{22} 
\end{array}\right), \; 0 \le t \le T,
$$
where, with abuse of notation, $\sigma_{11}, \sigma_{21} : [0,T] \times E \rightarrow \mathbb{R}$ are now (measurable) \textit{functions} of $(t,r(t))$, and $\sigma_{22} \in \mathbb{R}$ is a constant. 

Furthermore, we assume that the \textit{market price of risk} process $\lambda(\cdot)$ is given by $\lambda(t) = (\lambda_1(t,r(t)),\lambda_2)$, where, again with abuse of notation, $\lambda_1:[0,T] \times E \rightarrow \mathbb{R}$ is now a measurable function of $(t,r(t))$, and $\lambda_2 \in \mathbb{R}$ is a constant. 

Thus, the price processes $S_1$ and $S_2$ now have dynamics given by
\begin{align*}
    dS_1(t)/S_1(t) & = \left(r(t) + \lambda_1(t,r(t)) \sigma_{11}(t,r(t)) \right) \, dt + \sigma_{11}(t,r(t)) \, dW_1(t), \\
    dS_2(t)/S_2(t) &= \left(r(t) + \lambda_1(t,r(t)) \sigma_{21}(t,r(t)) + \lambda_2 \sigma_{22} \right) \, dt \\
    & \hspace{1em} + \sigma_{21}(t,r(t)) \, dW_1(t) + \sigma_{22} \, dW_2(t). 
\end{align*}
We think of $\lambda_1(\cdot,\cdot)$ and $\lambda_2$ as the market prices of interest rate risk and (pure) stock risk, respectively. Thus, $S_1$ is the price process of an interest rate derivative, e.g., a zero coupon bond, whereas we think of $S_2$ as the price process of a stock (or a mutual fund). For simplicity, we refer to them as the bond and the stock, respectively. The coefficients $\sigma_{11}$ and $\sigma_{21}$ are the volatilities of the bond and the stock, respectively, with respect to interest rate risk, and $\sigma_{22}$ is the volatility of the stock with respect to pure stock risk. In this way, the stock price process can be correlated with the bond price process. 

In this special case, we shall only allow investment strategies that are functions of the state variables $(t,r)$ (see Remark~\ref{rem_state_variables} below). So again, with abuse of notation, $\pi=(\pi_1, \pi_2) : [0,T] \times E \rightarrow \mathbb{R}^2$, is now assumed to be a measurable $\mathbb{R}^2$-valued mapping of $(t,r) \in [0,T] \times E$. Moreover, to narrow our search for martingale consumption, we only allow consumption strategies of the form (\ref{c_defined_by_a}), where we now further require that $Z(t) = a(t,r(t))$ for some continuous function $a:[0,T] \times E \rightarrow [0,\infty)$, which is strictly positive on $[0,T) \times E$ and fulfills 
$$
\int_0^T \frac{1}{a(t,r(t))} \, dt = \infty, \; \as,
$$
implying, in particular, that
\begin{equation} \label{a_zero_at_T}
    a(T,r) = 0, \; \forall r \in E.
\end{equation} 
The wealth dynamics (\ref{X_dynamics_with_c_as_factor}) can now be written as
\begin{align}
    dX(t) & = X(t) \, \left( r(t) -\frac{1}{a()}+ D^\pi_X() \lambda_1() \sigma_r() + \pi_2() \lambda_2 \sigma_{22} \right) \, dt  \nonumber \\
    & + X(t) D^\pi_X() \sigma_r() \, dW_1(t) + X(t) \pi_2() \sigma_{22} \, dW_2(t), \label{Wealth_dynamics_general_Specialized}
\end{align}
where
$$
D^\pi_X(t,r(t)) = \frac{\sum_{i=1}^2 \pi_i(t,r(t)) \sigma_{i1}(t,r(t))}{\sigma_r(t,r(t))}, \; 0 \le t \le T,
$$
denotes the sensitivity of $X$ to interest rate fluctuations, which depends on $\pi(\cdot)$, and where we have used the short-hand notation $()$ for $(t,r(t))$ for all functions of $(t,r(t))$. 

Now, we are interested in the dynamics of the consumption process $c$ under the restrictions on its form imposed above. If $a \in C^{1,2}([0,T]\times E)$, then Ito's lemma yields (recall (\ref{Z_dynamics}))
\begin{eqnarray} 
    \alpha_Z(t) & = & \frac{\partial a}{\partial t}() +\mathcal{A}_ta(), \label{alpha_Z} \\
    \sigma_Z(t) & = & \left(- \frac{\partial a}{\partial r}()\sigma_r(), \; 0 \right), \label{sigma_Z}
\end{eqnarray} 
where, once again, we have used the short-hand notation $()$ for $(t,r(t))$, and where $\mathcal{A}_t$ is the operator defined from (\ref{short_rate_general}) by
$$
\mathcal{A}_ta(t,r) = \frac{\partial a}{\partial r}(t,r) \mu(t,r) + \frac{\partial^2 a}{\partial r^2}(t,r)\frac{\sigma_r^2(t,r)}{2}, \; (t,r)\in (0,T)\times E.
$$
Inserting (\ref{alpha_Z}) and (\ref{sigma_Z}) into (\ref{c_dynamics_general}), and performing a few lines of (linear) algebra, we obtain
\begin{align}
dc(t) & = c(t) \alpha_c(t,r(t)) \, dt \label{c_drift_general}  \\ 
& \hspace{1em} + c(t) \left(D^\pi_X()-D_a()\right) \sigma_r() \, dW_1(t) + c(t) \pi_2() \sigma_{22} \, dW_2(t), \label{c_vol_general}
\end{align}
with $\alpha_c :(0,T) \times E \rightarrow \mathbb{R}$ given by
\begin{align} 
    \alpha_c(t,r) & = r+D^\pi_X()\lambda_1()\sigma_r()+\pi_2()\lambda_2 \sigma_{22} \nonumber \\
    & \hspace{1em} - \frac{1}{a()}\left(\frac{\partial a}{\partial t}()+\mathcal{A}_ta() -\frac{\partial a}{\partial r}() \sigma_r^2()\left(D^\pi_X()-D_a()\right) + 1\right) ,\label{c_drift}
\end{align}
where $()$ is short-hand notation for $(t,r)$, and $D_a(t,r)$ denotes (minus) the sensitivity of $a$ to interest rate fluctuations, i.e.,
$$
D_a(t,r) = -\frac{\partial a(t,r)/\partial r}{a(t,r)}, \; (t,r) \in [0,T)\times E. 
$$
We are now ready to state the main result in the stochastic interest rate case.
\begin{proposition} \label{prop_main}
Let $\pi$ be given. Assume that $a() \in C^{1,2}([0,T]\times E)$ satisfies the second order parabolic PDE
\begin{eqnarray*}
0 & = & \frac{\partial a}{\partial t}(t,r) +\mathcal{A}a(t,r) +1  - \frac{\partial a(t,r)}{\partial r} \sigma_r^2(t,r)\left(D^\pi_X(t,r)-D_a(t,r)\right)\\
& & - a(t,r)\left[r+D^\pi_X(t,r)\lambda_1(t,r)\sigma_r(t,r)+\pi_2(t,r)\lambda_2 \sigma_{22}\right],
\end{eqnarray*}
on $(0,T) \times E$, with boundary condition 
$$
a(T,r) = 0, \; \forall r \in E.
$$
Then the consumption process $c$ is a local martingale.

Moreover, if 
\begin{equation} \label{Novikov}
\mathrm{E}\left[\exp \left(\frac{1}{2}\int_0^T \sigma_c^2(t) \, dt\right)\right]<\infty,    
\end{equation}
then $c$ is a true martingale.
\end{proposition}
\textsc{Proof}. It immediately follows from (\ref{c_drift_general}) and (\ref{c_drift}) that if $a()$ satisfies the PDE stated in the proposition, then $c$ has zero drift, i.e., $\alpha_c(t,r) \equiv 0$, and therefore is a local martingale. The latter part follows from the Novikov condition.\qed % \hfill $\Box$

\vspace{1em}

\noindent The PDE for $a$ is only semilinear, due to the appearance of $D_a = -(\partial a/\partial r)/a$ in the coefficient for $\partial a/\partial r$. No known general closed-form solution exists. However, under certain conditions on $\mu$ and $\sigma_r$, existence of a solution can be established; see \cite{Friedman64} for details. 

The PDE generalizes the ODE (\ref{a_ODE_BS}) from the case with deterministic interest rate. The difference consists of the added term $(D^\pi_X\lambda_1\sigma_r)(t,r)$ in the coefficient for $-a$ and, of course, the terms involving $\partial a/\partial r$ and $\partial^2 a/\partial r^2$.

One could hope that for certain specific interest rate models the PDE would become tractable. However, getting rid of the $D_a$-term would require either that $a$ did not depend on $r$, which is unnatural in general, or that $D_a \equiv D_X$. We shall return to the latter possibility below. 

Since it might be considered a natural suggestion for $a(\cdot)$, we briefly consider the special case where it has the form 
\begin{equation} \label{a_form}
a(t,r) = B_{rg(t,\cdot)+h(t;\cdot)}(t) = \int_t^T e^{-\int_t^u rg(t,s)+h(t,s) \, ds} \, du, \; (t,r) \in [0,T]\times E,    
\end{equation}
for some deterministic functions $g,h\in C^{1,0}([0,T]^2)$ with $g(t,t) = 1, \; \forall t \in [0,T]$. It is well known that in certain simple models such as \cite{Vasicek77}, the expressions $\meanmeas{\int_t^Te^{-\int_t^u r(s) \, ds}\, du}{\mathrm{(t,r)}}$ and $\int_t^T e^{-\int_t^u \meanmeas{r(s)}{\mathrm{(t,r)}} \, ds}\, du$ (with obvious notation) both have this form. Note that $a$ is strictly positive on $[0,T) \times E$ and zero at $(T,\cdot)$, as required. 

By the Leibniz integral rule we can generalize (\ref{B_diff_eq}) to get
\begin{equation} \label{da_dt}
\frac{\partial a}{\partial t}(t,r) = \frac{dB_{rg(t,\cdot)+h(t;\cdot)}(t)}{dt}
= (r+h(t,t))a(t,r) - 1 - H_1(t,r) - H_2(t,r), 
\end{equation}
for $(t,r) \in (0,T)\times E$, where 
\begin{align}
H_1(t,r) &= \int_t^Te^{-\int_t^u rg(t,s)+h(t,s) \, ds}\int_t^u r\frac{\partial g}{\partial t} (t,s) \, ds \, du, \label{H1} \\
H_2(t,r) &= \int_t^Te^{-\int_t^u rg(t,s)+h(t,s) \, ds}\int_t^u \frac{\partial h}{\partial t} (t,s) \, ds \, du. \label{H2} 
\end{align} 
In this case the PDE in Proposition \ref{prop_main} simplifies to
\begin{align}
0 &= - H_1(t,r) - H_2(t,r) + \frac{\partial a(t,r)}{\partial r} \left[\mu(t,r)-\sigma^2(t,r)\left(D^\pi_X(t,r)-D_a(t,r)\right)\right] \nonumber \\
 & \hspace{1em} + \frac{\partial^2 a(t,r)}{\partial r^2}\frac{\sigma^2(t,r)}{2} - a(t,r)\left[\left(D^\pi_X\lambda_1\sigma_r\right)(t,r)+\pi_2(t,r)\lambda_2 \sigma_{22}-h(t,t)\right],\label{a_PDE_annuity_case}
\end{align}
and as noted above, the boundary condition is satisfied. However, the PDE is still semilinear and in general has no known solutions. In particular, the results from the simple model (\ref{f_3}) and the more general model, Proposition~\ref{PropSolutionWithDeterministicCoeffs}, do not hold when the interest rate is stochastic.

However, it may be possible to obtain solution approximations where the drift term is kept close to zero, and this may be relevant in practical pension product design. As mentioned in the introduction, there are pension products with a predetermined asset allocation scheme, where, e.g., the proportion invested in stocks depends only on time since retirement, and where the remaining proportion is invested in some mutual low-risk, bond-based fund, say. Since smooth consumption is a general objective in our approach, it would also be a natural objective for such products to keep the benefit volatility low. From (\ref{c_vol_general}) we see that the total volatility of $c$ at time $t \in (0,T)$ is given by
\begin{equation}
\sigma_c(t) :=  \left(\left(D^\pi_X()-D_a()\right)^2 \sigma_r^2()+\pi_2^2() \sigma_{22}^2 \right)^{1/2}, \; 0 < t < T,
\end{equation}
where, again, $() = (t,r(t))$. 

Ideally, we would like to keep both terms of $\sigma_c^2(t)$ low. Now, $D_a$ will typically be decreasing and converge to zero as $t \rightarrow T$. Different retirees may have different time horizons, and with predetermined schemes for $\pi_1$ and $\pi_2$, $D^\pi_X()-D_a()$ would typically be increasing (at least from some time point), leading to higher volatility towards the end. Typically (but depending on the parameters), this could be mitigated by letting $\pi_2$ be decreasing over time. 

We end this section by considering the case with a \textit{fixed} consumption rate, i.e., an annuity certain, where the investment strategy is the hedging strategy. The value at time $t\in [0,T]$ of a fixed unit consumption stream, given that $r(t)=r$, is (with obvious notation)
$$
\overline{a}(t,r) = \meanmeas{\int_t^Te^{-\int_t^u r(s) \, ds}\, du}{\mathrm{Q};\mathrm{(t,r)}}.
$$
With the initial wealth $x_0$ our agent can thus get a fixed consumption rate at the level $c_0 := x_0/ \overline{a}(0,r(0))$. Now, referring to the Feynman-Kac Theorem (\cite{KaratzasShreve91}, Theorem~5.7.6), we have, under certain regularity conditions,
\begin{eqnarray}
0 & = & \frac{\partial \overline{a}}{\partial t}(t,r) +1 - r\overline{a}(t,r) \nonumber \\
& & + \frac{\partial \overline{a}(t,r)}{\partial r} \left[\mu(t,r)+\lambda_1(t,r)\sigma_r(t,r)\right] + \frac{\partial^2 \overline{a}(t,r)}{\partial r^2}\frac{\sigma_r^2(t,r)}{2}. \label{ann_PDE}
\end{eqnarray} 
When all wealth is invested in the fixed rate annuity, i.e., $X(t) \equiv c_0 \overline{a}(t,r(t))$, we thus have from Ito's formula and (\ref{ann_PDE}) that
\begin{eqnarray*}
dX(t) & = & c_0 \left(r(t) \overline{a}(t,r(t))-1\right) \, dt - c_0 \frac{\partial \overline{a}(t,r(t))}{\partial r} \sigma_r(t,r(t)) \, dW^{\mathrm{Q}}_1(t) \\    
& = & X(t) r(t) \, dt - c_0 \, dt + X(t) D_{\overline{a}}(t,r(t)) \sigma_r(t,r(t))\, dW^{\mathrm{Q}}_1(t),
\end{eqnarray*}
where 
$$
D_{\overline{a}}(t,r) = -\frac{\partial \overline{a}(t,r)/\partial r}{\overline{a}(t,r)}, \; (t,r) \in [0,T)\times E. 
$$
Thus, comparing with the wealth dynamics (\ref{Wealth_dynamics_general_Specialized}) we see that the investment strategy for the fixed rate annuity is given by $\pi_2 \equiv 0$ and 
\be \label{D_pi_Ann}
D^{\pi}_X(t,r) = D_{\overline{a}}(t,r),
\ee
i.e., 
$$
\pi_1(t,r) = D_{\overline{a}}(t,r)\frac{\sigma_r(t,r)}{\sigma_1(t,r)}, \; (t,r) \in [0,T) \times E.
$$
With this investment strategy the term involving $\left(D^\pi_X(t,r)-D_a(t,r)\right)$ in the PDE of Proposition~\ref{prop_main} vanishes due to (\ref{D_pi_Ann}), and the PDE itself then simplifies to (\ref{ann_PDE}), i.e., $\overline{a}$ does indeed satisfy the PDE of Proposition~\ref{prop_main} if it satisfies (\ref{ann_PDE}). In certain simple and well-known affine term structure models such as the \cite{Vasicek77} model, $\overline{a}(\cdot)$ can be written in closed form, and one can then verify directly that $\overline{a}$ is sufficiently regular and thus satisfies (\ref{ann_PDE}) and the PDE of Proposition~\ref{prop_main}.

\begin{remark} \label{rem_state_variables}
    %\normalfont 
    In the special case in this section we have only allowed the (proportional) investment strategy and the wealth-to-consumption factor function to depend on the state variables $(t,r(t))$. For full generality, they should admittedly be allowed to depend on $X(t)$ as well. However, we have chosen this restriction for simplicity. Incorporating $X(t)$ would just lead to a more general PDE for the wealth-to-consumption factor as a function of $(t,r,x)$, for which we would still not have a closed-form solution. It is also worth noting that for a CRRA utility-optimizing agent, the optimal proportional investment allocation generally does not depend on the current wealth (under certain conditions on the time preference structure). \qed %\hfill $\Box$
\end{remark}

\section{The discrete-time case} \label{Sec:DiscreteTimeCase}
Since we have not been able to provide solutions in the general case for arbitrary investment strategies in continuous time, we briefly consider a discrete-time version of the problem. 

Thus, consider a discretization $0 = t_0 < t_1 < \ldots < t_n = T$ for some integer $n \ge 2$. Let $X_i$ denote the wealth at time $t_i$, $i = 0, \ldots, n$. For $i=1,\ldots,n$, we denote by $C_i$ and $R_i$ the amount consumed at time $t_i$, and the investment return in the $i$'th time interval $(t_{i-1},t_i]$, respectively (we thus assume that no consumption takes place at time $t_0$, which is of course just a convention). The wealth evolves according to 
\be \label{X_dynamics_discrete}
X_i = X_{i-1} (1+R_i)-C_i, \; i=1,\ldots,n.
\ee
The returns $R_i$ are assumed to be integrable random variables with values in $(-1,\infty)$, and we let $\mathcal{F}_i=\sigma(R_1,\ldots,R_i), \; i=1,\ldots,n$. Now, assume that $C_i = X_{i-1}(1+R_i)/a_i, \; i=1,\ldots,n$, for some adapted strictly positive process $(a_i)_{i=1,\ldots,n}$, and that $a_n \equiv 1$, so that the remaining wealth is consumed at time $n$. From (\ref{X_dynamics_discrete}) and the assumption about $C$, a few lines of simple algebra yields
$$
C_{i+1}= C_i(a_i-1)\frac{1+R_{i+1}}{a_{i+1}}, \; i = 1,\ldots,n-1.
$$
Thus, $C$ is a martingale if and only if $a$ satisfies the equation  
\be \label{eq_a_discrete}
\condmean{\frac{1+R_{i+1}}{a_{i+1}}}{\mathcal{F}_i} = \frac{1}{a_i-1}, 
\ee
$\forall i = 1,\ldots,n-1$. In the general case, (\ref{eq_a_discrete}) must be solved recursively, starting with $a_n\equiv 1$, and then 
$$
    a_i = 1+\frac{1}{\condmean{\frac{1+R_{i+1}}{a_{i+1}}}{\mathcal{F}_i}},\; i=n-1,\ldots,1.
$$
Unfortunately, there seems to be no nice formula for $a_i$ in general. On the contrary, note that for $i=1,\ldots,n-1$, $a_i$ is a function of $(R_1,\ldots,R_i)$, so we need the joint distribution of the returns to perform the calculations in each step. However, at least a solution exists in this case, as also shown by \cite{Fischer08}.  

It is well known that if the $R_i$ all represent a fixed interest rate, say $R_i \equiv r, \; i=1,\ldots,n$, for some $r>-1$, then $a$ given explicitly by the annuity formula
$$
a_i = 1+ \sum_{j=i+1}^n \frac{1}{(1+r)^{j-i}} = \frac{(1+r)-(1+r)^{i-n}}{r}, \; i = 1,\ldots,n-1,
$$
yields a consumption martingale (with $C_i = X_0(1+r)/a_1, \, i=1,\ldots,n$). A natural candidate for $a$ in the general case could be to let
\be \label{a_discrete_cand}
a_i = 1+\sum_{j=i+1}^n \prod_{k=i+1}^j \frac{1}{\condmean{1+R_k}{\mathcal{F}_i}}, \; i = 1,\ldots,n-1,
\ee
which can be viewed as a discrete-time version of (\ref{B}) with the expected future returns, given the information at time $t_i$,  replacing the function $f$, which determines the discounting factor in (\ref{B}). And in fact, if the returns $R_1,\ldots,R_n$ are mutually independent, one may verify (carefully!) that (\ref{eq_a_discrete}) is satisfied with $a$ given by (\ref{a_discrete_cand}) (and in this case, we can of course skip the conditioning in the formulas). However, this does \textit{not} hold in general when the returns are not mutually independent. 
\section{Conclusion} \label{Sec:Conclusion}
We briefly summarize our results, beginning with the continuous-time models of Sections~\ref{Sec:SimpleModels}, \ref{Sec:GeneralCase}, and \ref{Sec:StochInterestCase}, in the order of generality, i.e., first \ref{Sec:GeneralCase}, then \ref{Sec:StochInterestCase}, and then \ref{Sec:SimpleModels}. 

\vspace{1em}

\noindent The general model of Section~\ref{Sec:GeneralCase}:
\begin{itemize}
    \item For an arbitrary, given investment strategy, we have shown that an MCS is generally unique. However, the question of existence of an MCS in general is unsettled.
    \item For an arbitrary \textit{deterministic} investment strategy we have provided an explicit solution in the case where $r$ and $\alpha$ are also deterministic.
    \item For CRRA utility-optimizing agents we have shown that simultaneous martingale consumption can be achieved through a specific choice of time preference function if $r$ and $\lambda$ are deterministic (and even in general if the time preference function is replaced by a suitably chosen time preference process). 
\end{itemize}
Since the results from the general case of course apply to the special cases considered in Sections~\ref{Sec:SimpleModels} and \ref{Sec:StochInterestCase}, we only report the specific special results obtained in those cases below.

\vspace{1em}
\noindent The stochastic interest rate model of Section~\ref{Sec:StochInterestCase}:
\begin{itemize}
    \item For an arbitrary investment strategy, modeled as a function of $(t,r(t))$ we have provided a PDE for the wealth-to-consumption factor function $a()$, for which a solution would yield an MCS. In general, though, it is a semilinear PDE without known solutions. Under certain regularity conditions, a solution can be shown to exist.
\end{itemize}

\vspace{1em}
\noindent The simple model with deterministic coefficients of Section~\ref{Sec:SimpleModels}:
\begin{itemize}
    \item Any deterministic investment strategy allows for an explicit solution.
\end{itemize}

\vspace{1em}
\noindent The discrete-time model of Section~\ref{Sec:DiscreteTimeCase}:
\begin{itemize}
    \item A solution generally exists, but an explicit solution is generally only easy to provide if the returns in each time period are mutually independent.
\end{itemize}

As a final remark, we note the similarity between the results in the continuous- and discrete-time models: If the infinitesimally \textit{expected} return over $(t,t+dt)$, given $\mathcal{F}_t$, respectively, the \textit{expected} return over $(t_{i-1},t_i]$, given $\mathcal{F}_{i-1}$, is known at time 0 for all $t \in [0,T)$, resp.\ for all $i=1,\ldots,n$, i.e., does not depend on $\mathcal{F}_t$ resp.\ $\mathcal{F}_{i-1}$, then the natural solution candidate given by the wealth-to-consumption factor determined by the expected total return over the remaining period in fact leads to an MCS (recall (\ref{f_3}), Proposition~\ref{PropSolutionWithDeterministicCoeffs}, and (\ref{a_discrete_cand})). This is the case when the economy is not driven by random factors. In more realistic models, in particular models with stochastic interest rates, where the expected future return at some time point, given the evolution until then, actually depends on that evolution, then the natural solution candidate does not generally lead to an MCS.

\section*{Acknowledgments}
The author wants to thank Prof. Mogens Steffensen (Univ. Copenhagen) for early discussions of the approach suggested in this paper and for pointing out several relevant references. The author takes full responsibility for the contents of the paper.

\section*{Conflict of interest}
The author declares no conflict of interest.

\bibliographystyle{myapalike}
\bibliography{MC}

\end{document}